\documentclass[aps,eqsecnum,preprint,preprintnumbers,12pt,amsfonts]{revtex4}
\usepackage{epsfig,psfrag}

\usepackage{latexsym}
\usepackage{graphicx}
\usepackage{axodraw}

\newcommand{\be}{\begin{equation}}
\newcommand{\ee}{\end{equation}}
\newcommand{\ba}{\begin{eqnarray}}
\newcommand{\ea}{\end{eqnarray}}
\def\nn{\nonumber}

\def\olmass{\begin{picture}(10,10)(10,10)
\PhotonArc(15,10)(10,0,180)2 5
\Line(0,10)(30,10)
\end{picture}}

\def\tlscbf{\begin{picture}(10,10)(10,10)
\Photon(4,15)(26,15)2 4
\CArc(15,15)(10,0,180)
\CArc(15,15)(10,180,360)
\CArc(15,15)(12,0,180)
\CArc(15,15)(12,180,360)
\end{picture}}

\def\tlscfrph{\begin{picture}(10,10)(10,10) 
\DashLine(5,15)(25,15){1}                   
\CArc(15,15)(10,0,180)                      
\CArc(15,15)(10,180,360)                    
\end{picture}}                              

\def\olscbfphvvb{\begin{picture}(10,10)(10,10)   
\DashLine(5,15)(25,15){1}                       
\CArc(15,15)(12,0,180)                          
\CArc(15,15)(12,180,360)                        
\Vertex(15,15)2                                 
\put(13,18){\scriptsize 2}                      
\end{picture}}                                  

\def\tlscbfphnk{\begin{picture}(10,10)(10,10)    
\DashLine(5,15)(25,15){1}                        
\CArc(15,15)(10,0,180)                           
\CArc(15,15)(10,180,360)                         
\Vertex(15,5)2                                  
\Vertex(15,25)2 
\put(13,29){\scriptsize n}  
\put(13,-4){\scriptsize k}                     
\end{picture}}                                    

\def\tlscbfphnnkk{\begin{picture}(10,10)(10,10)  
\DashLine(5,15)(25,15){1}                      
\CArc(15,15)(10,0,180)                         
\CArc(15,15)(10,180,360)                       
\Vertex(15,5)2                                 
\Vertex(15,25)2                                
\put(8,29){\scriptsize n+1}                     
\put(10,-4){\scriptsize k-1}                     
\end{picture}} 
                                                                       
\def\tlscfr{\begin{picture}(10,10)(10,10)
\Photon(5,15)(25,15)2 4
\CArc(15,15)(10,0,180)
\CArc(15,15)(10,180,360)
\end{picture}}

\def\olscbf{\begin{picture}(10,10)(10,10)
\CArc(15,15)(10,0,180)
\CArc(15,15)(10,180,360)
\CArc(15,15)(12,0,180)
\CArc(15,15)(12,180,360)
\end{picture}}

\def\olscfr{\begin{picture}(10,10)(10,10)
\CArc(15,15)(10,0,180)
\CArc(15,15)(10,180,360)
\end{picture}}

\def\olscfrvdva{\begin{picture}(10,10)(10,10) 
\CArc(15,15)(10,0,180)                        
\CArc(15,15)(10,180,360)                      
\Vertex(15,5)2                                
\put(13,-4){\scriptsize 2}                    
\end{picture}}                                

\def\olscfrvtri{\begin{picture}(10,10)(10,10)
\CArc(15,15)(10,0,180)
\CArc(15,15)(10,180,360)
\Vertex(15,5)2
\put(13,-4){\scriptsize 3}
\end{picture}}

\def\olscbfvdva{\begin{picture}(10,10)(10,10)
\CArc(15,15)(10,0,180)
\CArc(15,15)(10,180,360)
\CArc(15,15)(12,0,180)
\CArc(15,15)(12,180,360)
\Vertex(15,5)2
\put(13,-4){\scriptsize 2}
\end{picture}}

\def\olscfrvst{\begin{picture}(10,10)(10,10)
\CArc(15,15)(10,0,180)
\CArc(15,15)(10,180,360)
\Vertex(15,5)2
\put(13,-4){\scriptsize 4}
\end{picture}}
                         
\def\olscfrvn{\begin{picture}(10,10)(10,10)
\CArc(15,15)(10,0,180) 
\CArc(15,15)(10,180,360) 
\Vertex(15,5)2
\put(13,-4){\scriptsize n}
\end{picture}}

\def\olscfrvk{\begin{picture}(10,10)(10,10)  
\CArc(15,15)(10,0,180)                       
\CArc(15,15)(10,180,360)                     
\Vertex(15,5)2                               
\put(13,-4){\scriptsize k}                   
\end{picture}}

\def\olscfrvnn{\begin{picture}(10,10)(10,10)
\CArc(15,15)(10,0,180) 
\CArc(15,15)(10,180,360) 
\Vertex(15,5)2
\put(8,-4){\scriptsize n+1}
\end{picture}}

\begin{document}

%\title{``Background Field Integration-by-Parts'' and Mass Renormalization in Two-Loop Heisenberg-Euler Effective Actions}
\title{``Background Field Integration-by-Parts'' and the Connection Between One-Loop and Two-Loop Heisenberg-Euler Effective Actions}

\author{Gerald V. Dunne\footnote{dunne@phys.uconn.edu} and Marek Kras\v nansk\'y\footnote{mkras@phys.uconn.edu}}

\affiliation{Department of Physics,
University of Connecticut,
Storrs, CT 06269-3046, USA}

\begin{abstract}
We develop integration-by-parts rules for diagrams involving massive scalar propagators in a constant background electromagnetic field, and use these to show that there is a simple diagrammatic interpretation of mass renormalization in the two-loop scalar QED 
Heisenberg-Euler effective action for a general constant background field. This explains why the square of a one-loop term appears in the renormalized two-loop Heisenberg-Euler effective action.
No integrals need be evaluated, and the explicit form of the background field propagators is not needed. This dramatically simplifies the computation of the renormalized two-loop effective action for scalar QED, and
generalizes a previous result obtained for self-dual background fields.
\end{abstract}

\maketitle

\section{Introduction}

Great progress has been made in recent years in computing higher-loop Feynman diagrams in quantum field theories \cite{smirnovbook,steinhauser,tarasov,kotikov,avdeev,baikov,laporta,anastasiou,glover,bern,smirnov,schroder}. A key ingredient of this program is the idea of  ``integration-by-parts" rules, and similar algebraic manipulations, which reduce diagrams to simpler forms before they need to be evaluated \cite{chetyrkin,bender,vladimirov}. Very recently, remarkable recursion formulas have been found relating amplitudes at different loop orders in ${\mathcal N}=4$ SYM \cite{bdks}. On the other hand, a complementary approach to studying such amplitudes is to consider their generating function, the effective action. There has also been some new progress in recent years in understanding the two-loop structure of Heisenberg-Euler effective actions, in QED \cite{dunneschubert,dgs}, super QED \cite{kuzenkoqed} and super Yang-Mills \cite{kuzenkosym}, with self-dual backgrounds. This effective action generates two-loop amplitudes with any number of external lines, and definite helicities, in the low momentum limit \cite{louise}. For QED in a self-dual background, a simple recursive relation between one-loop and two-loop was found \cite{sdloops,dunnekogan}, of a form analogous to the amplitude relations in \cite{bdks}.  The purpose of this paper is to establish a direct connection between the aforementioned advances in higher-loop amplitude computations and these advances in effective action computations. Specifically, we derive the renormalized two-loop scalar QED effective action using new ``integration-by-parts'' rules valid for massive scalar propagators {\it in a constant electromagnetic background field}.  We show that the identification of the square of a one-loop term in the fully renormalized two-loop scalar QED effective action has a natural algebraic origin that does not require evaluation of any integrals. The explicit form of the background field propagators is not needed; only the equation that they satisfy. This approach has the greatest potential for extending the two-loop results to higher loops. At the two-loop level it is considerably simpler than other direct evaluations of two-loop Heisenberg-Euler effective actions \cite{ritus,dittrich,csreview,schmidt,fliegner,kors,sato}. It also has the potential to make connection with Kreimer's Hopf algebra approach to renormalization of quantum field theory \cite{dirk}.

\section{Background field "integration-by-parts" rules}
\label{byparts}

The "integration-by-parts" method finds algebraic relations between diagrams at different loop order, without actually evaluating the diagrams \cite{chetyrkin}. The new relations we find in this paper are for bubble diagrams involving propagators in background fields. The approach can be motivated by similar ideas for bubble diagrams involving {\it free} propagators \cite{broadhurst,avdeev,baikov,schroder}. A simple example involving free propagators occurs in scalar QED, where the two-loop vacuum bubble diagram is proportional to the square of the one-loop bubble diagram: 
\ba 
\tlscfr \hskip .3cm =
\frac{e^2}{2}\left(\frac{d-1}{d-3}\right)\Bigl[ \hskip .3cm\mbox{\olscfr}\hskip .3cm  \Bigr]^2 \quad . 
\label{free1lsq} 
\ea
We work in $d$-dimensional Euclidean space with dimensional regularization of diagrams, and the solid line denotes a massive scalar propagator, and the wavy line denotes a Feynman gauge photon propagator.
The important observation is that equation (\ref{free1lsq}) can be derived by purely
algebraic means as follows. First, by simple manipulations (valid in dimensional regularization \cite{thooft}) of the integrand, it can be reduced to scalar diagrams: 
\ba
\tlscfr \hskip.3cm &=&\frac{e^2}{2} \int \frac{d^dp\, d^dq}{(2\pi)^{2d}}\, \frac{(p+q)^2}{(p-q)^2(p^2+m^2)(q^2+m^2)}\nn \\ \nn\\
&=& \frac{e^2}{2} \int \frac{d^dp\, d^dq}{(2\pi)^{2d}}\, \frac{\left[-(p-q)^2+2(p^2+m^2)+2(q^2+m^2)-4 m^2\right]}{(p-q)^2(p^2+m^2)(q^2+m^2)}\nn \\ \nn \\
&=& -\frac{e^2}{2}\Bigl[ \hskip .3cm \olscfr  \hskip .3cm \Bigr]^2
+ 2e^2 \int \frac{d^dp\, d^dq}{(2\pi)^{2d}}\, \frac{1}{(p-q)^2(p^2+m^2)}
-2 e^2 m^2  \Bigl[  \hskip .3cm \tlscfrph  \hskip .3cm\Bigr]\quad .
\label{freestep1} 
\ea 
Here the dotted line denotes a massless scalar
propagator. The first term has been written as the square of a
one-loop diagram. The second term vanishes as the integral over $q$ is zero. But the third term is apparently still two-loop.
However, using integration-by-parts manipulations in the following way, this two-loop
diagram can also be written as a square of a one-loop diagram.
We start from an identity valid in dimensional regularization:
\ba                                                      
0 &=& \int \frac{d^dp\,                                 
d^dq}{(2\pi)^{2d}}\,\frac{\partial}{\partial p_\mu}     
\left[\frac{(p-q)_\mu}{(p-q)^2} \frac{1}{(p^2+m^2)(q^2+m^2)} \right]       \nn \\    \nn \\                    
&=& \int \frac{d^dp\,                                                                                                   
d^dq}{(2\pi)^{2d}}\, \left[ \frac{d-2}{(p-q)^2} \frac{1}{(p^2+m^2)(q^2+m^2)} - 2 \frac{p(p-q)}{(p-q)^2}\frac{1}{(p^2+m^2)^2(q^2+m^2)}  \right]    \nn \\    \nn \\  
&=&(d-3)\Bigl[ \hskip .3cm \tlscfrph  \hskip .3cm\Bigr] - \Bigl[ \hskip .3cm \olscfrvdva  \hskip .3cm  \Bigr] \, \Bigl[\hskip .3cm \mbox{\olscfr} \hskip .3cm\Bigr] \quad .
\label{freestep2} 
\ea
Here we have used the simple identity, $2 p\cdot(p-q)=(p-q)^2+(p^2+m^2)-(q^2+m^2)$, and the notation of the solid dot with a number by a propagator indicates that that propagator is raised to that power.
Finally, the relation
\ba                                                                   
\olscfrvdva \quad = \frac{1}{2m^2} (2-d) \quad \olscfr  \qquad ,
\label{freestep3}
\ea      
follows from another integration-by-parts identity:
\ba
0=\int \frac{d^dp}{(2\pi)^d}\,      \frac{\partial}{\partial p_\mu}     
\left(\frac{p_\mu}{p^2+m^2} \right) \quad .
\ea                                                      
Combining (\ref{freestep2}) and (\ref{freestep3}) we find that the two-loop diagram appearing on the RHS of (\ref{freestep1}) is proportional to a one-loop diagram squared:
\ba
\tlscfrph  \hskip .3cm = - \frac{1}{2m^2} \frac{d-2}{d-3} \hskip .3cm \Bigl[ \hskip .3cm \olscfr  \hskip .3cm \Bigr]^2
\label{freemassless}
\ea
This then proves (\ref{free1lsq}).

We have found analogous algebraic manipulations for the two-loop
bubble diagrams for scalar QED in a general constant background
electromagnetic field. The presence of the background field has two
effects. First, it modifies the scalar propagator. In fact, our
manipulations do not rely on using the explicit form of the scalar
propagator in such a background (even though such an expression is
well known \cite{fock,nambu,schwinger}). Instead, we only use the equation satisfied by this
propagator, the background field Klein-Gordon equation, which in momentum space reads: 
\ba 
(p^2+m^2)G(p) =
1+\frac{e^2}{4} F_{\mu\alpha} F_{\nu\alpha} \frac{\partial^2
G(p)}{\partial p_\mu \partial p_\nu} \quad .
\label{kg}
\ea
This simplifies the analysis and has the greatest potential for extending the two-loop results to higher loops.

The second consequence of the background field is that it modifies the vertices, since $p_\mu \to p_\mu-i\frac{e}{2}\, F_{\mu\nu}\frac{\partial}{\partial p_\nu}$.
Therefore, the two-loop bubble diagram is
\ba
\tlscbf \hskip.3cm &=& \frac{e^2}{2} \int \frac{d^dp\, d^dq}{(2\pi)^{2d}}\, \frac{1}{(p-q)^2}\left\{ (p+q)^2G(p)G(q)-e^2 F_{\mu\alpha} F_{\nu\alpha} \frac{\partial G(p)}{\partial p_\mu}  \frac{\partial G(q)}{\partial q_\nu} \right\}\,\, .
\label{bfmanip}
\ea
Here we introduce the notation that the double line denotes the propagator in the background field.
By integration by parts, we can move both derivative operators (symmetrically) onto a single $G$ propagator, and then use the propagator Klein-Gordon equation (\ref{kg}) to yield
\ba
\tlscbf \hskip.3cm &=& \frac{e^2}{2} \int \frac{d^dp\, d^dq}{(2\pi)^{2d}}\, \frac{1}{(p-q)^2}\left\{ (p+q)^2+2(p^2+m^2)+2(q^2+m^2)\right\} G(p)G(q)\,\, .
\label{bfmanip2}
\ea

Motivated by the free-propagator identity (\ref{free1lsq}), we write $(p+q)^2$ as:
\ba                                                                                                 
\hskip -.7cm (p+q)^2=\left(\frac{d-1}{d-3}\right)(p-q)^2                           
 +2\left\{ (p^2+m^2)+(q^2+m^2)-2 m^2 -\left(\frac{d-2}{d-3}\right) (p-q)^2 \right\}  
\ea                                                                                                 
Therefore,
\ba
\tlscbf \hskip .3cm &=&\frac{e^2}{2}\left( \frac{d-1}{d-3}\right) \Big[\hskip .3cm \mbox{\olscbf}\hskip .3cm\Big]^2
\label{bf1lsq}\\
&+& e^2 \int \frac{d^dp\, d^dq}{(2\pi)^{2d}}\,\left\{ \frac{2}{(p-q)^2}\left[(p^2+m^2)+(q^2+m^2) - m^2\right] -\left(\frac{d-2}{d-3}\right)\right\} G(p)G(q)
\nn
\ea
Notice that in the absence of the background field, $G \to G_0$, the second term vanishes by (\ref{freemassless}), and so we recover the free scalar field result (\ref{free1lsq}).

The bare two-loop effective action is given by the difference of the two-loop bubble diagrams with and without the background field. Combining (\ref{free1lsq}) and (\ref{bf1lsq}) we obtain
\ba
 \Big[\hskip .3cm \mbox{\tlscbf}\hskip .3cm- \hskip .3cm\mbox{\tlscfr}\hskip .3cm\Big]&=& \frac{e^2}{2}\left( 
\frac{d-1}{d-3}\right) \left\{\Big[\hskip .3cm \mbox{\olscbf}\hskip .3cm\Big]^2- \Big[\hskip .3cm \mbox{\olscfr}\hskip .3cm 
\Big]^2 \right\}
\label{diff}\\
&&\hskip -3cm + e^2 \int \frac{d^dp\, d^dq}{(2\pi)^{2d}}\,\left\{ \frac{2}{(p-q)^2}\left[(p^2+m^2)+(q^2+m^2)-m^2\right] -
\left(\frac{d-2}{d-3}\right)\right\} G(p)G(q)
\nn
\ea
Now recall that in free scalar QED the one-loop mass  renormalization shift is given by a diagram that can also be  manipulated algebraically to the form                                    
\ba                                                                      
\delta m^2\equiv \left[\hskip                                                  
.3cm \olmass \hskip .3cm\right]_{p^2=-m^2}=e^2\left(                     
\frac{d-1}{d-3}\right) \hskip .3cm\olscfr\hskip .3cm        
\quad .                                                                  
\label{massshift}                                                        
\ea                                                                      
Note the same dimension dependent numerical coefficient in     
(\ref{free1lsq}) and (\ref{massshift}). 

Thus, consider the first two terms on the RHS of (\ref{diff}), and complete the square :
\ba
\Bigl[\hskip .3cm\olscbf\hskip .3cm\Bigr]^2-\Bigl[\hskip .3cm \olscfr \hskip .3cm\Bigr]^2 = \Bigl[ \hskip .3cm \olscbf
\hskip .3cm -\hskip .3cm \olscfr\hskip .3cm\Bigr]^2 +2  \Bigl[\hskip .3cm\olscfr\hskip .3cm\Bigr] \, \Bigl[ \hskip .3cm\olscbf\hskip .3cm -\hskip .3cm \olscfr\hskip .3cm \Bigr]
\label{complete}
\ea
The one-loop difference is, by definition, the derivative of the renormalized one-loop effective Lagrangian, up to an $O(F^2)$ charge renormalization term:
\ba
\Bigl[ \hskip .3cm \olscbf\hskip .3cm -\hskip .3cm \olscfr\hskip .3cm \Bigr] = -\frac{\partial {\mathcal L}^{(1)}_{\rm ren}}{\partial (m^2)} - \frac{e^2}{2d} F_{\mu\nu}F_{\mu\nu}
\Bigl[ \left( d-4 \right) \hskip .3cm \mbox{\olscfrvtri} \hskip .3cm  +4m^2 \hskip .3cm \mbox{\olscfrvst}\hskip .3cm \Bigr] \quad .
\label{der1l} 
\ea 
Thus, the two-loop difference (\ref{diff}) can be written as
\ba
\Big[\hskip .3cm \mbox{\tlscbf}\hskip .3cm- \hskip
.3cm\mbox{\tlscfr}\hskip .3cm\Big]&=&\frac{e^2}{2}\left(
\frac{d-1}{d-3}\right) \Big[\hskip .3cm \mbox{\olscbf}\hskip .3cm-
\hskip .3cm \mbox{\olscfr}\hskip .3cm \Big]^2- \delta m^2\,
\frac{\partial {\mathcal L}^{(1)}_{\rm ren}}{\partial (m^2)}+O(F^2)
\label{massrenorm}\\
&&\hskip -3cm + e^2 \int \frac{d^dp\, d^dq}{(2\pi)^{2d}}\,\left\{
\frac{2}{(p-q)^2}\left[(p^2+m^2)+(q^2+m^2)-m^2\right]
-\left(\frac{d-2}{d-3}\right)\right\} G(p)G(q) \nn 
\ea
The first term on the RHS is the square of a one-loop term, analogous to the RHS of (\ref{free1lsq}), and moreover is finite in $d=4$. The second term is just the mass renormalization term. The remaining terms, discussed below, all vanish of course in the absence of a background field.

So the first main observation of this paper is that the mass renormalization part of the two-loop effective Lagrangian can be separated out from the bare two-loop effective Lagrangian by a series of straightforward algebraic steps, with no need to evaluate any integrals. This is in dramatic contrast
to direct evaluations using the explicit propertime integral representations of the background field propagators \cite{ritus,dittrich,csreview,schmidt,fliegner,kors,sato,dunneschubert,dgs} where the mass renormalization is isolated through divergences of complicated double-integrals over the two proper-time parameters (one for each propagator). We also observe that this procedure of mass renormalization automatically identifies a term in the renormalized two-loop effective action which is the square of a one-loop term, the first term on the RHS of (\ref{massrenorm}). This generalizes the relation (\ref{free1lsq}) for free propagators to the case of propagators in a background field. Next we turn to the remaining terms in (\ref{massrenorm}).

After removing the mass renormalization term,
 $\delta m^2\, \frac{\partial {\mathcal L}^{(1)}}{\partial (m^2)}$, the only possible remaining divergence in the two-loop effective Lagrangian is associated with charge renormalization,
  which must arise in a term proportional to the bare Maxwell Lagrangian $F^2$. Therefore, we can neglect the $O(F^2)$ term coming from (\ref{der1l}), and any $O(F^2)$ terms coming from the last integral in (\ref{massrenorm}):
  \ba
\hskip -.5cm e^2 \int \frac{d^dp\, d^dq}{(2\pi)^{2d}}\,\left\{ \frac{2\left[(p^2+m^2)+(q^2+m^2)\right]}{(p-q)^2}
  - \left( \frac{2m^2}{(p-q)^2} + \frac{d-2}{d-3}\right)\right\}\, G(p)\,G(q)\quad .
\label{remainder} 
\ea 
It is helpful to split this remainder into two pieces. Applying the Klein-Gordon equation (\ref{kg}) to the first part we obtain
\ba                             
\Sigma_1 &\equiv& 2 e^2 \int \frac{d^dp\, d^dq}{(2\pi)^{2d}}\,\frac{\left[(p^2+m^2)+(q^2+m^2)\right] }{(p-q)^2}G(p)G(q)\nn\\
&= &  e^4 F_{\mu\alpha} F_{\nu \alpha} \int \frac{d^dp\,
d^dq}{(2\pi)^{2d}}\,\left( \frac{\partial^2}{\partial p_\mu \partial p_\nu}\frac{1}{(p-q)^2}\right) G(p)G(q) 
\label{massren} 
\ea 
The divergent $F^2$ part of (\ref{massren}) arises when the
propagators inside the integral are replaced with free ones, in
which case this term yields a term proportional to the Maxwell
Lagrangian 
\be 
e^4 F_{\mu\nu}F_{\mu\nu} \frac{2(4-d)}{d} \quad
\olscbfphvvb  \hskip .3cm
= \frac{e^4}{2m^4} F_{\mu\nu}F_{\mu\nu} \frac{(d-4)(d-2)}{d(5-d)}\, \Bigl[ \hskip .3cm
\olscfr\hskip .3cm \Bigr]^2\quad .
 \ee
Thus, 
\ba
\Sigma_1&=&e^4 F_{\mu\alpha} F_{\nu \alpha} \int \frac{d^dp\,
d^dq}{(2\pi)^{2d}}\,\left( \frac{\partial^2}{\partial p_\mu \partial
p_\nu}\frac{1}{(p-q)^2} \right) \left[G(p)G(q)-G_0(p) G_0(q)
\right] \nn\\
&&\hskip 1cm +\frac{e^4}{2m^4} F_{\mu\nu}F_{\mu\nu} \frac{(d-4)(d-2)}{d(5-d)}\, \Bigl[ \hskip .3cm
\olscfr\hskip .3cm \Bigr]^2\quad .
\label{sigma1-finite}
\ea
The first term is manifestly finite in $d=4$, and is $O(F^4)$.

The second part of the remainder term (\ref{remainder}) 
\ba 
\Sigma_2 = e^2 \int \frac{d^dp\, 
d^dq}{(2\pi)^{2d}}\,\left\{ 
\frac{2m^2}{(p-q)^2}+ \frac{d-2}{d-3} \right\} G(p)G(q) \quad ,
\label{sigma2} 
\ea
vanishes when the propagators are replaced by
free ones, by virtue of (\ref{freemassless}). In fact, $\Sigma_2$ is completely finite in $d=4$, even with the background field propagators. To see this,
we use the Klein-Gordon equation (\ref{kg}) to expand the full scalar propagator in a background field, $G(p)$, as an expansion in powers of the field-strength tensor $F$ and the free scalar propagator $G_0(p)=1/(p^2+m^2)$:
\ba                                                                                                                                
G(p)&=&  G_0(p)+ \frac{e^2}{4} G_0(p) F_{\mu\alpha} F_{\nu\alpha}\frac{\partial^2 G_0(p)}{\partial p_\mu \partial p_\nu} \nn\\
&&\hskip 1cm + \left(   
       \frac{e^2}{4} \right)^2 G_0(p) F_{\mu\alpha} F_{\nu\alpha}F_{\varrho\beta}F_{\sigma\beta} \frac{\partial^2}{\partial p_\mu  
         \partial p_\nu} \left[G_0(p) \frac{\partial^2 G_0(p)}{\partial p_\varrho \partial p_\sigma} \right]  +\dots
         \label{G(p)}                                                                                                                                        
\ea                              
Then using an integration-by-parts identity 
(\ref{app2}) derived in the Appendix, we obtain: 
\ba 
\Sigma_2 =  
&-&\frac{m^2e^4}{d-3} \ F_{\mu\alpha} F_{\nu\alpha}  \int \frac{d^dp\, 
d^dq}{(2\pi)^{2d}}\,\left( \frac{\partial^2}{\partial p_\mu \partial 
p_\nu}\frac{1}{(p-q)^2} \right) G_0^2(p) G_0(q)                                    \label{sigma2-finite}  \\
&-&\frac{m^2e^6}{4(d-3)} \ F_{\mu\alpha} F_{\nu\alpha} \ F_{\varrho\beta} F_{\sigma\beta}  \int \frac{d^dp\, 
d^dq}{(2\pi)^{2d}}\,\left( \frac{\partial^2}{\partial p_\mu \partial 
p_\nu}\frac{1}{(p-q)^2} \right) G_0^2(p) \ G_0(q) \frac{\partial^2 G(q)}{\partial q_\varrho \partial q_\sigma} \nn \\
&+&e^2\,\left(\frac{e^2}{4}\right)^2 F_{\mu\alpha} F_{\nu\alpha} \ F_{\varrho\beta} F_{\sigma\beta}  \int \frac{d^dp\, 
d^dq}{(2\pi)^{2d}}\, \left\{ \frac{2m^2}{(p-q)^2}+\frac{d-2}{d-3}\right\}
G_0(p) \frac{\partial^2 G(p)}{\partial p_\mu \partial p_\nu} \ G_0(q) \frac{\partial^2 G(q)}{\partial q_\varrho \partial q_\sigma}
\nn
\ea 
Due to the isotropy of $G_0(p)$, the first integral in (\ref{sigma2-finite}) is proportional to $\delta_{\mu\nu}$. 
Because this integral is finite, we can set $d=4$ and 
use 
\ba
\square\,  \frac{1}{(p-q)^2}=-4\pi^2\delta^4(p-q)
\label{delta}
\ea
Thus $\Sigma_2$ becomes
\ba
\Sigma_2=\frac{m^2e^4}{16 \pi^2} \ F_{\mu\nu}F_{\mu\nu} \,\Bigl[ \quad \olscfrvtri \quad \Bigr] +O(F^4)
\label{proof4} 
\ea
By the renormalizability of scalar QED, the $O(F^4)$ must be finite, and this can be confirmed by simple power-counting arguments for the integrals appearing in (\ref{sigma2-finite}).

We therefore find the finite renormalized two-loop effective Lagrangian as:
\ba
\Big[\hskip .3cm \mbox{\tlscbf}\hskip .3cm- \hskip
.3cm\mbox{\tlscfr}\hskip .3cm\Big]_{\rm ren} &=&\frac{e^2}{2}\left(
\frac{d-1}{d-3}\right) \Big[\hskip .3cm \mbox{\olscbf}\hskip .3cm-
\hskip .3cm \mbox{\olscfr}\hskip .3cm \Big]^2\nn\\
&&+ e^4 F_{\mu\alpha} F_{\nu \alpha} \int \frac{d^dp\,
d^dq}{(2\pi)^{2d}}\,\left( \frac{\partial^2}{\partial p_\mu \partial
p_\nu}\frac{1}{(p-q)^2} \right) \left[G(p)G(q)-G_0(p) G_0(q)
\right]\nn\\
&\hskip -3cm -\hskip+3cm &\hskip -3cm \frac{m^2e^6}{4(d-3)} \ F_{\mu\alpha} F_{\nu\alpha} \ F_{\varrho\beta} F_{\sigma\beta}  \int \frac{d^dp\, 
d^dq}{(2\pi)^{2d}}\,\left( \frac{\partial^2}{\partial p_\mu \partial 
p_\nu}\frac{1}{(p-q)^2} \right) G_0^2(p) \ G_0(q) \frac{\partial^2 G(q)}{\partial q_\varrho \partial q_\sigma} \nn \\
&\hskip -3cm +\hskip+3cm &\hskip -3cm e^2\,\left(\frac{e^2}{4}\right)^2 F_{\mu\alpha} F_{\nu\alpha} \ F_{\varrho\beta} F_{\sigma\beta}  \int \frac{d^dp\, 
d^dq}{(2\pi)^{2d}}\, \left\{ \frac{2m^2}{(p-q)^2}+\frac{d-2}{d-3}\right\}
G_0(p) \frac{\partial^2 G(p)}{\partial p_\mu \partial p_\nu} \ G_0(q) \frac{\partial^2 G(q)}{\partial q_\varrho \partial q_\sigma}\nn\\ \nn\\
\label{answer}
\ea
Each term on the RHS is finite in $d=4$, and the first term is the square of a one-loop object.

\section{Self-dual background field}
\label{SDcase}

So far the discussion is valid for a general constant background field strength $F_{\mu\nu}$. In \cite{sdloops,dunnekogan} it was shown that an even simpler expression than (\ref{answer}) is obtained if the background field is self-dual. As explained in \cite{sdloops,dunnekogan}, the $d$-dimensional generalization of ``self-dual" is the condition that $F_{\mu\alpha}F_{\nu\alpha}=f^2\delta_{\mu\nu}$. 
This dramatically simplifies the form of both $\Sigma_1$ in (\ref{sigma1-finite}) and $\Sigma_2$ in (\ref{sigma2-finite}). For example, using (\ref{delta}) we see that  (\ref{sigma1-finite}) becomes
 \ba
 \Sigma_1&=&- \ \frac{e^4 f^2}{4\pi^2} \Big[ \hskip .4cm \mbox{\olscbfvdva}\hskip .3cm -\hskip .3cm\mbox{\olscfrvdva}\hskip .3cm\Big] +\frac{e^4 f^2}{2m^4}  \frac{(d-4)(d-2)}{(5-d)}\, \Bigl[ \hskip .3cm
\olscfr\hskip .3cm \Bigr]^2 +O(d-4)\quad .
 \label{sigma1SD}
 \ea
Furthermore, for a self-dual field $\Sigma_2$ and $\Sigma_1$ are connected by an algebraic identity:
\ba
\Sigma_2=\frac{1}{2} \ \frac{d-4}{d-3} \ \Sigma_1
\label{identity}
\ea
The factor $d-4$ in front of $\Sigma_1$ in (\ref{identity}) cancels the $\frac{1}{d-4}$ divergence in the second term of  (\ref{sigma1SD}), and
$\Sigma_2$ becomes finite in $d=4$, 
and contains only integrals over free propagators.
  
Therefore, the fully renormalized two-loop effective Lagrangian in a self-dual background in $d=4$ can be written as
 \ba                                                                                                                               
 \hskip -.7cm \Big[\hskip .3cm \mbox{\tlscbf}\hskip .3cm- \hskip .3cm\mbox{\tlscfr}\hskip .3cm\Big]_{\rm ren} 
&=& \frac{3 e^2}{2} \Big[\hskip .3cm \mbox{\olscbf}\hskip .3cm -\hskip .3cm \mbox{\olscfr}\hskip .3cm \Big]^2  
-\frac{e^4 f^2}{4 \pi^2}\Big[ \hskip .4cm \mbox{\olscbfvdva}\hskip .3cm -\hskip .3cm\mbox{\olscfrvdva}\hskip .3cm\Big]       
\label{sdcase}                                                                                                                                                                                                                                                                                                    
 \ea                                                                                                                               
This is precisely the result found in \cite{sdloops,dunnekogan}, namely that the two-loop renormalized effective Lagrangian is written entirely in terms of one-loop objects, and that this result can be obtained without evaluating any integrals. We now see that this is a special case of the more general result (\ref{answer}), which shows similarly that the first term on the RHS, which is the square of a one-loop term, appears   naturally as a result of the mass renormalization of the two-loop effective Lagrangian, and separates algebraically without doing any integrals. Furthermore, having separated the $O(F^2)$ charge renormalization terms, the expression (\ref{answer}) is manifestly finite.

\section{Conclusions}
\label{conclusions}

To conclude, we have developed algebraic ``integration-by-parts'' rules for vacuum diagrams involving massive scalar propagators in constant background electromagnetic fields. This leads directly to a simple implementation of mass renormalization in the two-loop Heisenberg-Euler effective Lagrangian; this approach is much more direct than the mass renormalization identification in earlier work \cite{ritus,dittrich,csreview,schmidt,fliegner,kors,sato,dunneschubert,dgs}, and so is a promising candidate for extending the two-loop to one-loop relation to higher loops. This result generalizes the result (\ref{sdcase}) of \cite{dunneschubert,sdloops,dunnekogan}, where it was shown that the renormalized two-loop effective Lagrangian contains two one-loop components: the first is the square of  the one-loop trace of the propagator, and the second is the one-loop trace of the square of the propagator. This present paper shows that the first of these one-loop objects is generic, for any background field, while the second arises due to special properties of the self-dual background considered in \cite{sdloops,dunnekogan}. The background field integration-by-parts technique can clearly be generalized to spinor or supersymmetric propagators, and to nonabelian theories. Some related ideas using expansions of background field propagators in coordinate space to isolate divergences of diagrams were used in \cite{bornsen} to compute the three-loop $\beta$-function in Yang-Mills theory.

\section*{Acknowledgments}
We thank the US DOE for support through grant DE-FG02-92ER40716, and GD thanks Dirk Kreimer for helpful comments.

\appendix
\section{Integration by parts}
\label{appendix}

In this Appendix we derive an identity that is used in analyzing the integral $\Sigma_2$ defined in (\ref{sigma2}). A generalization of (\ref{freemassless}) can be derived in a similar way from an identity valid in dimensional 
regularization:
\ba 
0 &=& \int \frac{d^dp\,
d^dq}{(2\pi)^{2d}}\,\frac{\partial}{\partial p_\mu}
\left[\frac{(p-q)_\mu}{(p-q)^2} \ G^n_0(p) F(q) \right] 
\label{app1}
\ea 
Here $G^n_0(p)$ is a free scalar propagator raised to the n-th power,
and $F(q)$ is an arbitrary function, which could for example be taken equal to the background field scalar propagator $G(q)$. The n-th power of $G_0(p)$ could arise, for example, in the perturbative expansion (\ref{G(p)}) of the background field propagator $G(p)$. After applying the derivative
with respect to $p_\mu$ we obtain:
\ba
0 &=& \int \frac{d^dp\,
d^dq}{(2\pi)^{2d}}\, \left[ \frac{d-2}{(p-q)^2} G_0^n(p) F(q) - 2n \frac{p\cdot (p-q)}{(p-q)^2}G_0^{n+1}(p) F(q)  \right] \quad .\nn
\ea
Using $2 p\cdot (p-q)=(p-q)^2+(p^2+m^2)-(q^2+m^2)$, and the integral of $G_0^{n+1}$, which can be expressed using the identity: 
\ba 
\olscfrvnn \quad = \frac{1}{2m^2} \frac{2n-d}{n} \quad \olscfrvn  \quad ,\nn 
\ea 
we obtain the result:
 \ba
\int \frac{d^dp\, d^dq}{(2\pi)^{2d}}\,\left\{ \frac{2m^2}{(p-q)^2}+\frac{d-2}{d-3}\right\} G^n_0(p) F(q) &=& \frac{\left( d-4 \right) \left( n-1 \right)}{\left( d-3 \right) \left( d-2-n \right)}\int \frac{d^dp\, d^dq}{(2\pi)^{2d}}\, G^n_0(p) F(q) -\nn \\
&\hskip -1cm-\hskip +1cm&\hskip -1cm \frac{2n\, m^2}{d-2-n}\int \frac{d^dp\, d^dq}{(2\pi)^{2d}}\, \frac{1}{{\left( p-q\right)}^2} G^{n+1}_0(p) \frac{1}{G_0(q)} F(q) \nn \\ \label{app2}
\ea
If we choose $F(q)$ to be the $k^{\rm th}$ power of the free scalar propagator, $G_0^k(q)$, the identity (\ref{app2}) becomes:
\ba
2m^2 \hskip .3cm  \tlscbfphnk   \hskip .3cm   + \frac{d-2}{d-3} \hskip .3cm \olscfrvn \hskip .6cm  \olscfrvk \hskip .3cm 
= \frac{n-1}{d-2-n} \ \frac{d-4}{d-3} \hskip .3cm \olscfrvn \hskip .6cm  \olscfrvk \hskip .3cm  -
\frac{2n \ m^2}{d-2-n}  \hskip .3cm \tlscbfphnnkk  \hskip .3cm
\ea
For $k=1$, the two-loop integral on the right-hand side vanishes. If also $n=1$ the whole right-hand side turns to be zero and we
recover (\ref{freemassless}).


\begin{thebibliography}{01234}

\bibitem{smirnovbook}
  V.~A.~Smirnov, {\it Evaluating Feynman Integrals}, 
  Springer tracts in modern physics {\bf 211}, (Springer, Berlin, 2004).
  
\bibitem{steinhauser}
  M.~Steinhauser,
  ``Results and techniques of multi-loop calculations,''
  Phys.\ Rept.\  {\bf 364}, 247 (2002)
  [arXiv:hep-ph/0201075].
  %%CITATION = HEP-PH 0201075;%%
  
  \bibitem{tarasov}
  O.~V.~Tarasov,
  ``Reduction of Feynman graph amplitudes to a minimal set of basic
  integrals,''
  Acta Phys.\ Polon.\ B {\bf 29}, 2655 (1998)
  [arXiv:hep-ph/9812250].
  %%CITATION = HEP-PH 9812250;%%
  
    \bibitem{kotikov}
  A.~V.~Kotikov,
  ``Differential Equations Method: New Technique For Massive Feynman Diagrams
  Calculation,''
  Phys.\ Lett.\ B {\bf 254}, 158 (1991);
  %%CITATION = PHLTA,B254,158;%%
  ``Some methods for the evaluation of complicated Feynman integrals,''
  arXiv:hep-ph/0112347.
  %%CITATION = HEP-PH 0112347;%%
  
    \bibitem{avdeev}
  L.~V.~Avdeev,
  ``Recurrence Relations for Three-Loop Prototypes of Bubble Diagrams with a
  Mass,''
  Comput.\ Phys.\ Commun.\  {\bf 98}, 15 (1996)
  [arXiv:hep-ph/9512442];
  %%CITATION = HEP-PH 9512442;%%
  L.~V.~Avdeev, J.~Fleischer, M.~Y.~Kalmykov and M.~N.~Tentyukov,
  ``Towards automatic analytic evaluation of diagrams with masses,''
  Comput.\ Phys.\ Commun.\  {\bf 107}, 155 (1997)
  [arXiv:hep-ph/9710222].
  %%CITATION = HEP-PH 9710222;%%
  
   \bibitem{baikov}
  P.~A.~Baikov,
  ``Explicit solutions of the 3--loop vacuum integral recurrence relations,''
  Phys.\ Lett.\ B {\bf 385}, 404 (1996)
  [arXiv:hep-ph/9603267];
  %%CITATION = HEP-PH 9603267;%%
  ``Advanced methods of multi-loop integrals calculations: Status and
  perspectives,''
  Nucl.\ Phys.\ Proc.\ Suppl.\  {\bf 116}, 378 (2003);
  %%CITATION = NUPHZ,116,378;%%
   P.~A.~Baikov and M.~Steinhauser,
  ``Three-loop vacuum integrals in FORM and REDUCE,''
  Comput.\ Phys.\ Commun.\  {\bf 115}, 161 (1998)
  [arXiv:hep-ph/9802429].
  %%CITATION = HEP-PH 9802429;%%
  
  \bibitem{laporta}
  S.~Laporta,
  ``High-precision calculation of multi-loop Feynman integrals by  difference
  equations,''
  Int.\ J.\ Mod.\ Phys.\ A {\bf 15}, 5087 (2000)
  [arXiv:hep-ph/0102033].
  %%CITATION = HEP-PH 0102033;%%

  
  \bibitem{anastasiou}
  C.~Anastasiou and A.~Lazopoulos,
  ``Automatic integral reduction for higher order perturbative  calculations,''
  JHEP {\bf 0407}, 046 (2004)
  [arXiv:hep-ph/0404258].
  %%CITATION = HEP-PH 0404258;%%
  

\bibitem{glover}
  E.~W.~N.~Glover,
   ``Progress in NNLO calculations for scattering processes,''
  %
  Nucl.\ Phys.\ Proc.\ Suppl.\  {\bf 116}, 3 (2003)
  [arXiv:hep-ph/0211412].
  %%CITATION = HEP-PH 0211412;%%

%\cite{Bern:2002wt}
\bibitem{bern}
  Z.~Bern,
   ``Recent progress in perturbative quantum field theory. ((U)) ((W)),''
  %
  Nucl.\ Phys.\ Proc.\ Suppl.\  {\bf 117}, 260 (2003)
  [arXiv:hep-ph/0212406].
  %%CITATION = HEP-PH 0212406;%%
  
    
  \bibitem{smirnov}
  V.~A.~Smirnov and M.~Steinhauser,
  ``Solving recurrence relations for multi-loop Feynman integrals,''
  Nucl.\ Phys.\ B {\bf 672}, 199 (2003)
  [arXiv:hep-ph/0307088].
  %%CITATION = HEP-PH 0307088;%%
  
 
  
  \bibitem{schroder}
  Y.~Schroder,
  ``Automatic reduction of four-loop bubbles,''
  Nucl.\ Phys.\ Proc.\ Suppl.\  {\bf 116}, 402 (2003)
  [arXiv:hep-ph/0211288].
  %%CITATION = HEP-PH 0211288;%%
  
   \bibitem{chetyrkin}
  K.~G.~Chetyrkin and F.~V.~Tkachov,
  ``Integration By Parts: The Algorithm To Calculate Beta Functions In 4
  Loops,''
  Nucl.\ Phys.\ B {\bf 192}, 159 (1981);
  %%CITATION = NUPHA,B192,159;%%
  ``Infrared R Operation And Ultraviolet Counterterms In The Ms Scheme,''
  Phys.\ Lett.\ B {\bf 114}, 340 (1982);
  %%CITATION = PHLTA,B114,340;%%
  K.~G.~Chetyrkin, M.~Faisst, C.~Sturm and M.~Tentyukov,
  ``$\epsilon$-finite basis of master integrals for the integration-by-parts method,''
  arXiv:hep-ph/0601165.
  %%CITATION = HEP-PH 0601165;%%
  %%Cited 1 time in SPIRES-HEP
  
    %\cite{Bender:1976pw}
\bibitem{bender}
  C.~M.~Bender, R.~W.~Keener and R.~E.~Zippel,
   ``New Approach To The Calculation Of $F_{(1)}(\alpha)$ In Massless Quantum
   Electrodynamics,''
  %
  Phys.\ Rev.\ D {\bf 15}, 1572 (1977).
  %%CITATION = PHRVA,D15,1572;%%
  
  %\cite{Vladimirov:1979zm}
\bibitem{vladimirov}
  A.~A.~Vladimirov,
   ``Method For Computing Renormalization Group Functions In Dimensional
   Renormalization Scheme,''
  %
  Theor.\ Math.\ Phys.\  {\bf 43}, 417 (1980)
  [Teor.\ Mat.\ Fiz.\  {\bf 43}, 210 (1980)].
  %%CITATION = TMPHA,43,417;%%

\bibitem{bdks}
  C.~Anastasiou, Z.~Bern, L.~J.~Dixon and D.~A.~Kosower,
   ``Planar amplitudes in maximally supersymmetric Yang-Mills theory,''
  Phys.\ Rev.\ Lett.\  {\bf 91}, 251602 (2003)
  [arXiv:hep-th/0309040];
  %%CITATION = HEP-TH 0309040;%%
  Z.~Bern, L.~J.~Dixon and V.~A.~Smirnov,
  ``Iteration of planar amplitudes in maximally supersymmetric Yang-Mills
  theory at three loops and beyond,''
  Phys.\ Rev.\ D {\bf 72}, 085001 (2005)
  [arXiv:hep-th/0505205].
  %%CITATION = HEP-TH 0505205;%%
  
  %\cite{Dunne:2001pp}
\bibitem{dunneschubert}
  G.~V.~Dunne and C.~Schubert,
   ``Closed-form two-loop Euler-Heisenberg Lagrangian in a self-dual
   background,''
  %
  Phys.\ Lett.\ B {\bf 526}, 55 (2002)
  [arXiv:hep-th/0111134];
  %%CITATION = HEP-TH 0111134;%%
   ``Two-loop self-dual Euler-Heisenberg Lagrangians. I: Real part and  helicity
   amplitudes,''
  %
  JHEP {\bf 0208}, 053 (2002)
  [arXiv:hep-th/0205004];
  %%CITATION = HEP-TH 0205004;%%
   ``Two-loop self-dual Euler-Heisenberg Lagrangians. II: Imaginary part  and
   Borel analysis,''
  %
  JHEP {\bf 0206}, 042 (2002)
  [arXiv:hep-th/0205005].
  %%CITATION = HEP-TH 0205005;%%
  %%Cited 10 times in SPIRES-HEP
  
    %\cite{Dunne:2002ta}
\bibitem{dgs}
  G.~V.~Dunne, H.~Gies and C.~Schubert,
   ``Zero modes, beta functions and IR/UV interplay in higher-loop QED,''
  %
  JHEP {\bf 0211}, 032 (2002)
  [arXiv:hep-th/0210240].
  %%CITATION = HEP-TH 0210240;%%



%\cite{Kuzenko:2003qg}
\bibitem{kuzenkoqed}
  S.~M.~Kuzenko and I.~N.~McArthur,
   ``Low-energy dynamics in N = 2 super QED: Two-loop approximation,''
  %
  JHEP {\bf 0310}, 029 (2003)
  [arXiv:hep-th/0308136].
  %%CITATION = HEP-TH 0308136;%%

%\cite{Kuzenko:2004yd}
\bibitem{kuzenkosym}
  S.~M.~Kuzenko and I.~N.~McArthur,
   ``Relaxed super self-duality and effective action,''
  %
  Phys.\ Lett.\ B {\bf 591}, 304 (2004)
  [arXiv:hep-th/0403082].
  %%CITATION = HEP-TH 0403082;%%
  
  \bibitem{louise}
  L.~C.~Martin, C.~Schubert and V.~M.~Villanueva Sandoval,
  ``On the low-energy limit of the QED N-photon amplitudes,''
  Nucl.\ Phys.\ B {\bf 668}, 335 (2003)
  [arXiv:hep-th/0301022].
  %%CITATION = HEP-TH 0301022;%%
  
  \bibitem{sdloops}
G.~V.~Dunne,
``Two-loop diagrammatics in a self-dual background,''
JHEP {\bf 0402}, 013 (2004)
[arXiv:hep-th/0311167].
%%CITATION = HEP-TH 0311167;%%

\bibitem{dunnekogan}
  G.~V.~Dunne,
 ``Heisenberg-Euler effective Lagrangians: Basics and extensions,''  in Ian Kogan Memorial Collection, {\it From Fields to Strings: Circumnavigating Theoretical Physics}, Vol. I, M.~Shifman (ed.) et al, (World Scientific, 2005)  [arXiv:hep-th/0406216].
  %%CITATION = HEP-TH 0406216;%%



  
  
  \bibitem{ritus}
  V.~I.~Ritus,
   ``On The Relation Between The Quantum Electrodynamics Of An Intense Field And
   The Quantum Electrodynamics At Small Distances,''
  %
  Zh.\ Eksp.\ Teor.\ Fiz.\  {\bf 73}, 807 (1977);
  %%CITATION = ZETFA,73,807;%%
``The Lagrangian Function of an Intense Electromagnetic
Field'', in {\it Proc. Lebedev Phys. Inst.} Vol. {\bf 168}, {\it
Issues in Intense-field Quantum Electrodynamics}, V. I. Ginzburg,
ed., (Nova Science Pub., NY 1987).


%\cite{Dittrich:1985yb}
\bibitem{dittrich}
  W.~Dittrich and M.~Reuter,
   ``Effective Lagrangians In Quantum Electrodynamics,''
  %
  Lect.\ Notes Phys.\  {\bf 220}, 1 (1985).
  %%CITATION = LNPHA,220,1;%%

  
  %\cite{Schubert:2001he}
\bibitem{csreview}
  C.~Schubert,
   ``Perturbative quantum field theory in the string-inspired formalism,''
  %
  Phys.\ Rept.\  {\bf 355}, 73 (2001)
  [arXiv:hep-th/0101036].
  %%CITATION = HEP-TH 0101036;%%

  
  %\cite{Schmidt:1993rk}
\bibitem{schmidt}
  M.~G.~Schmidt and C.~Schubert,
   ``On the calculation of effective actions by string methods,''
  %
  Phys.\ Lett.\ B {\bf 318}, 438 (1993)
  [arXiv:hep-th/9309055];
  %%CITATION = HEP-TH 9309055;%%
   ``Worldline Green functions for multiloop diagrams,''
  %
  Phys.\ Lett.\ B {\bf 331}, 69 (1994)
  [arXiv:hep-th/9403158];
  %%CITATION = HEP-TH 9403158;%%
   ``Multiloop calculations in the string inspired formalism: The Single spinor
   loop in QED,''
  %
  Phys.\ Rev.\ D {\bf 53}, 2150 (1996)
  [arXiv:hep-th/9410100].
  %%CITATION = HEP-TH 9410100;%%



  
  %\cite{Fliegner:1997ra}
\bibitem{fliegner}
  D.~Fliegner, M.~Reuter, M.~G.~Schmidt and C.~Schubert,
   ``Two-loop Euler-Heisenberg Lagrangian in dimensional regularization,''
  %
  Theor.\ Math.\ Phys.\  {\bf 113}, 1442 (1997)
  [Teor.\ Mat.\ Fiz.\  {\bf 113}, 289 (1997)]
  [arXiv:hep-th/9704194].
  %%CITATION = HEP-TH 9704194;%%
 
 %\cite{Kors:1998ew}
\bibitem{kors}
  B.~Kors and M.~G.~Schmidt,
   ``The effective two-loop Euler-Heisenberg action for scalar and spinor  QED
   in a general constant background field,''
  %
  Eur.\ Phys.\ J.\ C {\bf 6}, 175 (1999)
  [arXiv:hep-th/9803144].
  %%CITATION = HEP-TH 9803144;%%

%\cite{Sato:1998sf}
\bibitem{sato}
  H.~T.~Sato and M.~G.~Schmidt,
   ``World-line approach to the Bern-Kosower formalism in two-loop  Yang-Mills
   theory,''
  %
  Nucl.\ Phys.\ B {\bf 560}, 551 (1999)
  [arXiv:hep-th/9812229].
  %%CITATION = HEP-TH 9812229;%%
  
  %\cite{Kreimer:1997dp}
\bibitem{dirk}
  D.~Kreimer,
   ``On the Hopf algebra structure of perturbative quantum field theories,''
  %
  Adv.\ Theor.\ Math.\ Phys.\  {\bf 2}, 303 (1998)
  [arXiv:q-alg/9707029];
  %%CITATION = Q-ALG 9707029;%%
  A.~Connes and D.~Kreimer,
   ``Renormalization in quantum field theory and the Riemann-Hilbert  problem.
   I: The Hopf algebra structure of graphs and the main theorem,''
  %
  Commun.\ Math.\ Phys.\  {\bf 210}, 249 (2000)
  [arXiv:hep-th/9912092].
  %%CITATION = HEP-TH 9912092;%%
  
  \bibitem{broadhurst}
  D.~J.~Broadhurst,
  ``Three loop on-shell charge renormalization without integration: Lambda-MS
  (QED) to four loops,''
  Z.\ Phys.\ C {\bf 54}, 599 (1992).
  %%CITATION = ZEPYA,C54,599;%%
  
  \bibitem{fock}
  V.~Fock,
   ``Proper Time In Classical And Quantum Mechanics,''
  %
  Phys.\ Z.\ Sowjetunion {\bf 12}, 404 (1937).
  %%CITATION = PHZSA,12,404;%%


\bibitem{nambu}
  Y.~Nambu,
   ``The Use Of The Proper Time In Quantum Electrodynamics,''
  %
  Prog.\ Theor.\ Phys.\  {\bf 5}, 82 (1950).
  %%CITATION = PTPKA,5,82;%%

\bibitem{schwinger}
  J.~S.~Schwinger,
  ``On Gauge Invariance And Vacuum Polarization,''
  Phys.\ Rev.\  {\bf 82}, 664 (1951).
  %%CITATION = PHRVA,82,664;%%


 \bibitem{thooft}
  G.~'t Hooft and M.~J.~G.~Veltman,
  ``Regularization And Renormalization Of Gauge Fields,''
  Nucl.\ Phys.\ B {\bf 44}, 189 (1972).
  %%CITATION = NUPHA,B44,189;%%



%\cite{Bornsen:2002hh}
\bibitem{bornsen}
  J.~P.~Bornsen and A.~E.~M.~van de Ven,
   ``Three-loop Yang-Mills beta-function via the covariant background field
   method,''
  %
  Nucl.\ Phys.\ B {\bf 657}, 257 (2003)
  [arXiv:hep-th/0211246].
  %%CITATION = HEP-TH 0211246;%%


 










\end{thebibliography}
\end{document}